\documentclass[usenatbib, referee]{mn2e}
\usepackage{graphicx}
\usepackage{ulem}

\usepackage{epsfig}
\usepackage{color}
\usepackage{amsmath}	
\usepackage{mathtools}
\usepackage{threeparttable}
\usepackage{booktabs}
\title[He-enrichment during nova outbursts]{Helium enrichment during classical nova outbursts}

\author[Y. Guo, C. Wu and B. Wang]{
Yunlang Guo, $^{\rm 1,2,3}$\thanks{E-mail:yunlang@ynao.ac.cn}
Chengyuan Wu$^{\rm 4}$ and
Bo Wang$^{\rm 1,2,3}$\thanks{E-mail:wangbo@ynao.ac.cn}
\\
$^{1}$Yunnan Observatories, Chinese Academy of Sciences, Kunming 650216, China\\
$^{2}$Key Laboratory for the Structure and Evolution of Celestial Objects, Chinese Academy of Sciences, Kunming 650216, China\\
$^{3}$University of Chinese Academy of Sciences, Beijing 100049, China\\
$^{4}$Physics Department and Tsinghua Center for Astrophysics (THCA), Tsinghua University, Beijing 100084, China
}

\date{Accepted XXX. Received YYY; in original form ZZZ}

\pubyear{2021}

\begin{document}
\label{firstpage}
\pagerange{\pageref{firstpage}--\pageref{lastpage}}
\maketitle

\begin{abstract}
Nova outbursts play an important role in the chemical evolution of galaxies, especially they are the main source of synthetic
$^{13}\rm C$, $^{15}\rm N$, $^{17}\rm O$ and some radioactive isotopes like $^{22}\rm Na$ and $^{26}\rm Al$.
The enrichment of He in nova ejecta indicates that 
the accreted material may mix with the He-shell (He-mixing).
The purpose of this work is to investigate how the He-mixing affects the nova outbursts in a systematic way.
We evolved a series of accreting WD models,
and found that the mass fraction of H and He in nova ejecta can be influenced by different He-mixing fractions significantly.
We also found that both the nova cycle duration and ejected mass increase with the He-mixing fractions.
Meanwhile, 
the nuclear energy generation rate of $p$-$p$ chains decreases with the He-mixing fraction during the nova outbursts,
whereas the CNO-cycle increases.
The present work can reproduce the chemical abundances in the ejecta of some representative novae, such as GQ Mus, ASASSN-18fv, HR Del, T Aur and V443 Sct.
This implies that the He-mixing process cannot be neglected when studying nova outbursts.
This study also develops a He-mixing meter (i.e. $\rm He/H$) that can be used to estimate the He-mixing fraction in classical nova systems.

\end{abstract}


\section{Introduction}

Classical novae are stellar explosions that occur on the surface of a white dwarf (WD) in binary systems
(e.g. Jos{\'e} 2016; Starrfield et al. 1972, 2020).
The WD accretes material from a main-sequence companion through the inner Lagrangian point when the donor fills its Roche-lobe,
triggering a thermonuclear runaway (TNR) in the accreted envelope.
Observationally, novae are mainly concentrated in the disk and bulge of the Galaxy, and the Galactic nova rate is $50_{-23}^{+31}$ per year (see Shafter 2017).

Theoretically, it is still uncertain whether the WD in nova systems can accumulate mass through successive outbursts. 
It has been claimed that the carbon-oxygen white dwarf (CO WD) in nova systems could increase its mass close to the Chandrasekhar limit ($M_{\rm Ch}$), 
indicating that nova is a potential candidate for the formation of type Ia supernovae (e.g. Hillebrandt \& Niemeyer 2000;
Wang \& Han 2012; Starrfield et al. 2020).
The final outcome of WDs could be an accretion-induced collapse event if the accretor is an oxygen-neon white dwarf (ONe WD) (e.g. Tauris et al. 2013; Wang \& Liu 2020). 
However, on the other hand,
most studies suggested that the WD in nova systems cannot increase its mass
(e.g. Nomoto et al. 2007; Hillman et al. 2016).
Yaron et al. (2005) suggested that most WDs would decrease their mass due to the H-shell flashes.
It has been argued that WDs may loss their mass since some of core matter is dredged up during the continuous outbursts
(e.g. MacDonald 1984; Starrfield et al. 2000).
In addition, He-shell under the H-shell may also undergo a TNR, causing the WD to reduce its mass (see Idan, Shaviv \& Shaviv 2013).

High abundance levels of WD material have been detected in nova ejecta, e.g. C, N, O, Ne, Na, Mg, and Al (e.g. Livio \& Truran
1994; Gehrz et al. 1998).
The enhanced metal abundances are strong evidences for mixing between the material of WD and the accreted matter. Therefore, the compositions of WDs can be identified by the spectroscopic studies, that is, CO and ONe.
However, the mixing process during accretion has not been fully understood so far.
Some mechanisms for explaining mixing process have been proposed.
For one-dimensional models, there are diffusion-induced convection (e.g. Prialnik \& Kovetz 1984; Fujimoto \& Iben 1992), shear mixing (e.g. Durisen 1977; Kippenhahn \& Thomas 1978; MacDonald 1983; Kutter \& Sparks 1987; Fujimoto 1988), convective overshoot-induced flame propagation (see, e.g. Woosley 1986).
For multidimensional models, there are mixing by resonant gravity waves (e.g. Rosner et al. 2001; Alexakis et al. 2004),
and Kelvin-Helmholtz instabilities (e.g. Glasner \& Livne 1995; Glasner et al. 2012; Casanova et al. 2010, 2016), etc.

Novae can provide about $0.3\%$ of the interstellar medium in the Galaxy, which is the main source of the nuclides like $^{13}\rm C$, $^{15}\rm N$, $^{17}\rm O$ (see, e.g. Jos{\'e} \& Hernanz 1998).
The radioactive isotopes $^{7}\rm Be$, $^{22}\rm Na$, $^{26}\rm Al$ can also be synthesized during nova outbursts (e.g.
Jos{\'e} \& Hernanz 1998; Starrfield et al. 1998; Denissenkov et al. 2014).
Therefore, novae are noteworthy contributors to the Galactic chemical evolution.
In addition, the elemental abundances of nova ejecta can reveal the properties of outbursts.
Downen et al. (2013) suggested that some elemental abundance ratios can be the most useful ``nova thermometers'', 
e.g. $\rm N/O$, $\rm N/Al$, $\rm O/S$, $\rm S/Al$, $\rm O/Na$, $\rm Na/Al$, $\rm O/P$ and $\rm P/Al$. 
These ratios are strongly dependent on the peak temperature during nova outbursts.
Moreover, a high N/O ratio can be produced in the nova outbursts with a massive WD (e.g. Politano et al. 1995; Jos{\'e} \& Hernanz 1998). 
 
It has been suggested that the enrichment of CNO or Ne in nova ejecta is caused by the mixing between the accreted layer and the outermost shells of the underlying WD.
\begin{figure}
	\centering\includegraphics[width=\columnwidth*3/5]{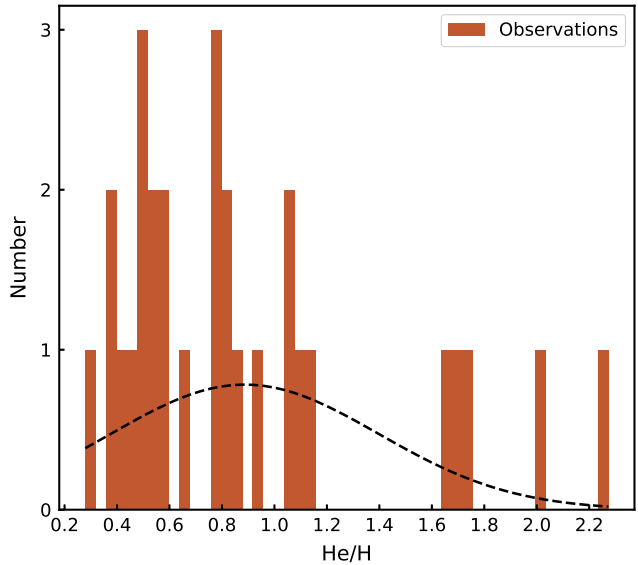}
	\caption{Distribution of the abundance (mass fraction) ratio of He to H ($\rm He/H$) in nova ejecta, in which the dashed line is fitted with the normal distribution. The observation data are taken from Gehrz et al. (1998).}
	\label{fig:ele1}
\end{figure}
However, significant enrichments of He are also considered to be a general characteristic in the ejecta of classical novae (e.g. Truran \& Livio 1986; Fujimoto \& Iben 1992; Starrfield et al. 1998).
Fig.\,1 shows the statistics of the abundance (mass fraction) ratio of He to H ($\rm He/H$) in nova ejecta, in which the data are taken from Gehrz et al. (1998).
From this figure, we can see that the peak of $\rm He/H$ is about 0.9 by mass fraction, and some even exceed 1.
This is because the He layer from the residual H-burning can be mixed with the accreted material (e.g. Starrfield et al. 1998; Jos{\'e} \& Hernanz 1998).
The He-enrichment is also observed in recurrent novae that contain a massive WD with a high accretion rate.
Particularly, the value of $\rm He/H$ in U Sco is suggested to be in the range of 2-4 by number (e.g. Webbink et al. 1987; Evans et al. 2001). 
By assuming that the outer He-shell is mixed with the accreted H layer, Glasner et al. (2012) recently suggested that the total amount of He mixing fraction is about 0.2.

Currently, it is a long-standing issue to reproduce the observed ejecta mass by theoretical simulations when studying nova outbursts (e.g. Kovetz \& Prialnik 1997; Jos{\'e} \& Hernanz 1998; Starrfield, Iliadis \& Hix 2016).
It has been generally believed that the ejected mass increases with the decrease of the WD mass, accretion rate, core temperature and metallicity (e.g. Yaron et al. 2005; Jos{\'e} et al. 2007; Chen et al. 2019).
It is worth noting that the ejected mass can also be increased by the He-mixing (see, e.g. Starrfield et al. 1998).
The purpose of this paper is to investigate the influence of He-mixing on the nova outbursts systematically.
In Sect. 2, we presented the basic assumptions and methods for numerical simulations.
The numerical results are shown in Sect. 3.
Finally, discussions and a summary are given in Sect. 4.
\section{Numerical Methods}
We used the stellar evolution code Modules for Experiments in Stellar Astrophysics (MESA, version 10398; see Paxton et al. 2011, 2013, 2015, 2018) to simulate the evolution accreting WDs.
Previous studies suggested that nova outbursts are mainly influenced by the WD mass ($M_{\rm WD}$), mass-accretion rate ($\dot M_{\rm acc}$), initial luminosity and the composition of matter accreted (e.g. Yaron et al. 2005; Wolf et al. 2013; Jos{\'e} 2016;
Rukeya et al. 2017; Wang 2018). 
In this work,
we created a series of WD models with different masses,
and cool them down until $10^{-2}\,\rm L_\odot$ to use as our initial models.

The classical novae are modeled by simulating the mass-accretion process onto the WDs.
Following the studies of Politano et al. (1995), the accreted material from the companion is mixed with the underlying WD matter (or mixed with He) by artificially setting a given fraction,
which is known as the pre-mixed model (e.g. Jos{\'e} \& Hernanz 1998; Denissenkov et al. 2014). 
In addition, we set the matter from companion as the solar abundances,
and the default OPAL opacity is used (e.g. Iglesias \& Rogers 1993, 1996).
The nuclear network in our simulations includes 68 isotopes from $^{1}\rm H-^{58}\rm Fe$, involving 402 nuclear reactions.
Moreover, the mixing-length parameter is set to be 2.0, and we do not consider the convective overshooting and rotation in the present work.

We performed a large number of calculations by using the super-Eddington wind as the mass loss mechanism during nova outbursts, which assumes that mass ejection is driven by the energy above the Eddington luminosity (e.g. Denissenkov et al. 2013, 2017).
The Eddington luminosity ($L_{\rm Edd}$) and the wind mass-loss rate ($\dot M$) are given by:
\begin{equation}
L_{\rm Edd} = \frac{4\pi GcM_{\rm WD}}{\tau},
\end{equation}
\begin{equation}
\dot M = \frac {2\eta_{\rm Edd}(L-L_{\rm Edd})}{\upsilon_{\rm esc}^2},
\end{equation} 	
where $c$ and $\tau$ are the vacuum speed of light and Rosseland mean opacity, $G$ and $\upsilon_{\rm esp}$ are the gravitational constant and escape velocity, $\eta_{\rm Edd}$ and $L$ are the super-Eddington wind factor and luminosity of the WD surface, respectively.
\begin{figure}
	\centering\includegraphics[width=\columnwidth*3/5]{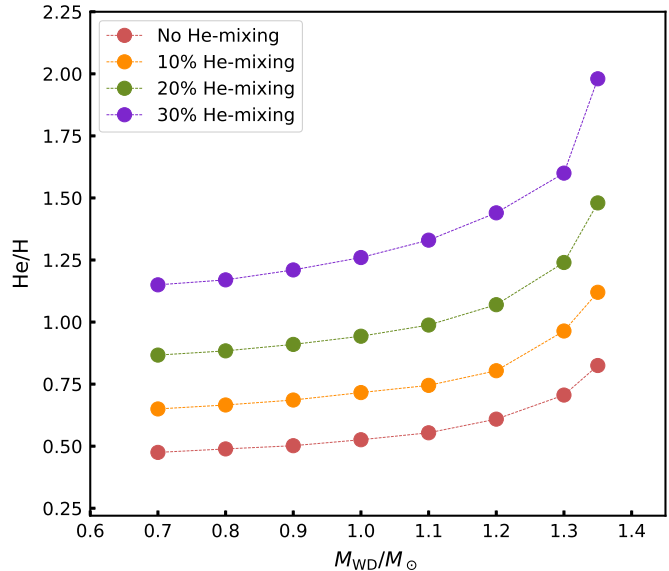}
	\caption{$\rm He/H$ changes with the He-mixing fractions ($f_{\rm He}$) for various initial WD masses ($M_{\rm WD}$), where we set the accretion rate to be $2 \times 10^{-9}\,\rm M_\odot \rm yr^{-1}$, and the initial WD mass to be $0.7-1.35\,\rm M_\odot$.
	The red, yellow, green and purple lines refer to the model without He-mixing, 10\%, 20\% and 30\% of $f_{\rm He}$, respectively.}
	\label{fig:com_herah5}
\end{figure}

\section{Results}
\subsection{The influence of the He-mixing}
We performed a large number of calculations for the nova models, 
in which we set the initial WD masses ($M_{\rm WD}$) to be $0.7-1.35\,\rm M_\odot$,
and the accretion rate to be $2 \times 10^{-9}\,\rm M_\odot \rm yr^{-1}$.
We mixed the accreted material with a given mass fraction of the He-shell,
and the He-mixing fraction ($f_{\rm He}$) is set to be $\leq 30\%$.
The mass fraction of H and He in the accreted material are calculated by: 
$X_{\rm H} = (1-f_{\rm He}) \times 0.7$ and 
$X_{\rm He} = 0.28 + 0.72 \times f_{\rm He}$, respectively.
In this part, we just explore the effects of He-mixing on the nova outbursts,
and do not involve the WD-mixing.

Fig.\,2 shows $\rm He/H$ changing with the initial WD masses for various He-mixing fractions.
The value of $\rm He/H$ increases significantly with the WD mass for a given He-mixing fraction when the WD mass is larger than $1.1\,\rm M_\odot$,
whereas $\rm He/H$ changes only slightly in the WD masses range of $0.7-1.1\,\rm M_\odot$.
This is because the maximum temperature increases more rapidly when $M_{\rm WD}\textgreater1.1\,\rm M_\odot$,
resulting in more H to burn into He (for more discussions, see Sect. 4).
We also note that $\rm He/H$ increases with the He-mixing level for a given WD mass.

The nova cycle duration ($D$) is defined as the time interval between two successive nova outbursts.
Fig.\,3 presents $D$ and the ejected mass ($M_{\rm ej}$) changing with WD masses for different He-mixing levels.
We note that both $D$ and $M_{\rm ej}$ decrease with the WD masses, but increase with the He-mixing levels.
Less accreted mass is expected in nova outbursts for massive WDs due to the high gravitational acceleration
(e.g. Yaron et al. 2005; Wolf et al. 2013; Hillman et al. 2016; Wu et al. 2017; Wang 2018; Guo et al. 2020).
The opacity of the envelope matter would be decreased if He is mixed in the accreted material,
leading to that the energy generated by the nuclear reaction and gravitation is rapidly radiated from the WD surface.
As a result, WDs need longer time to accrete the material required for nova outbursts, 
resulting in a more massive accreted layer and ejected mass.
It is worth noting that the ejected mass is only increased by a small amount in the He-mixing nova model.
Our simulations are not enough to explain the discrepancy, that is,
the observed ejecta mass is an order of magnitude or higher than that of the simulations.
Thus, the He-mixing is not expected to play an important role for the mass increase of nova ejecta.
\begin{figure}
	\centering\includegraphics[width=\columnwidth*3/5]{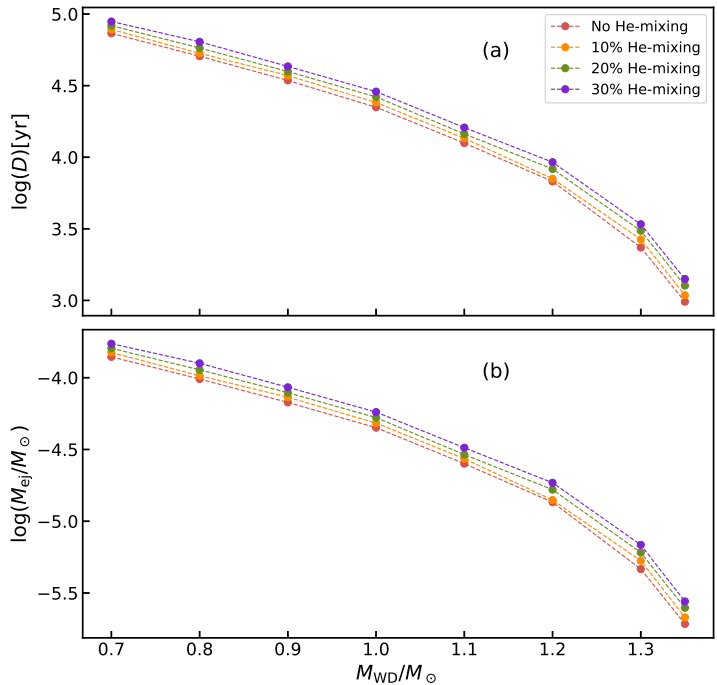}
	\caption{The influence of the WD mass and He-mixing fraction on the nova outburst characteristics. Panel (a): the results of the nova cycle duration $D$. Panel (b): the results of the ejected mass. 
	The red, yellow, green and purple lines represent the model without He-mixing, 10\%, 20\% and 30\% of $f_{\rm He}$, respectively.
	In the model without He-mixing, the accreted material is pure solar component.}
	\label{fig:gr}
\end{figure}

Fig.\,4 shows the comparison of nuclear energy generation rate of the $p$-$p$ chains and the CNO-cycle for different He-mixing degrees,
where we set the initial WD mass to be $1\,\rm M_\odot$.
During the nova outbursts,
the nuclear energy generation rate of the $p$-$p$ chains in the He-mixing model would decrease but the CNO-cycle increases.
This is because the less abundance of H in the He-mixing model,
leading to the decrease of the nuclear energy generation rate of $p$-$p$ chains (see Starrfield et al. 1998).
A longer accretion phase is required in the He-mixing model (see Fig.\,3a),
resulting in a massive accreted layer and a higher peak temperature.
Meanwhile, the nuclear energy generation rate of the CNO-cycle is sensitive to the temperature during nova outbursts
($T \textgreater 10^7\rm K$, $\epsilon_{\rm nuc} \sim T^{16-18}$),
and higher temperatures in the He-mixing models will lead to an increase in energy production.
Note that the nuclear energy generation rate of the CNO-cycle during accretion phase will decrease in the He-mixing model,
which is caused by the decrease in H abundance.

Fig.\,5 shows the comparison of luminosity evolution during a single nova outburst in different He-mixing levels,
where we set the initial WD mass to be $1.1\,\rm M_\odot$.
The peak of luminosity increases with the He-mixing level.
This is because more energy released by nuclear reaction in the He-mixing model,
which is consistent with that of Fig.\,4.
In the He-mixing model, the luminosity after the peak drops faster than that without He-mixing.

\begin{figure}
	\centering\includegraphics[width=\columnwidth*3/5]{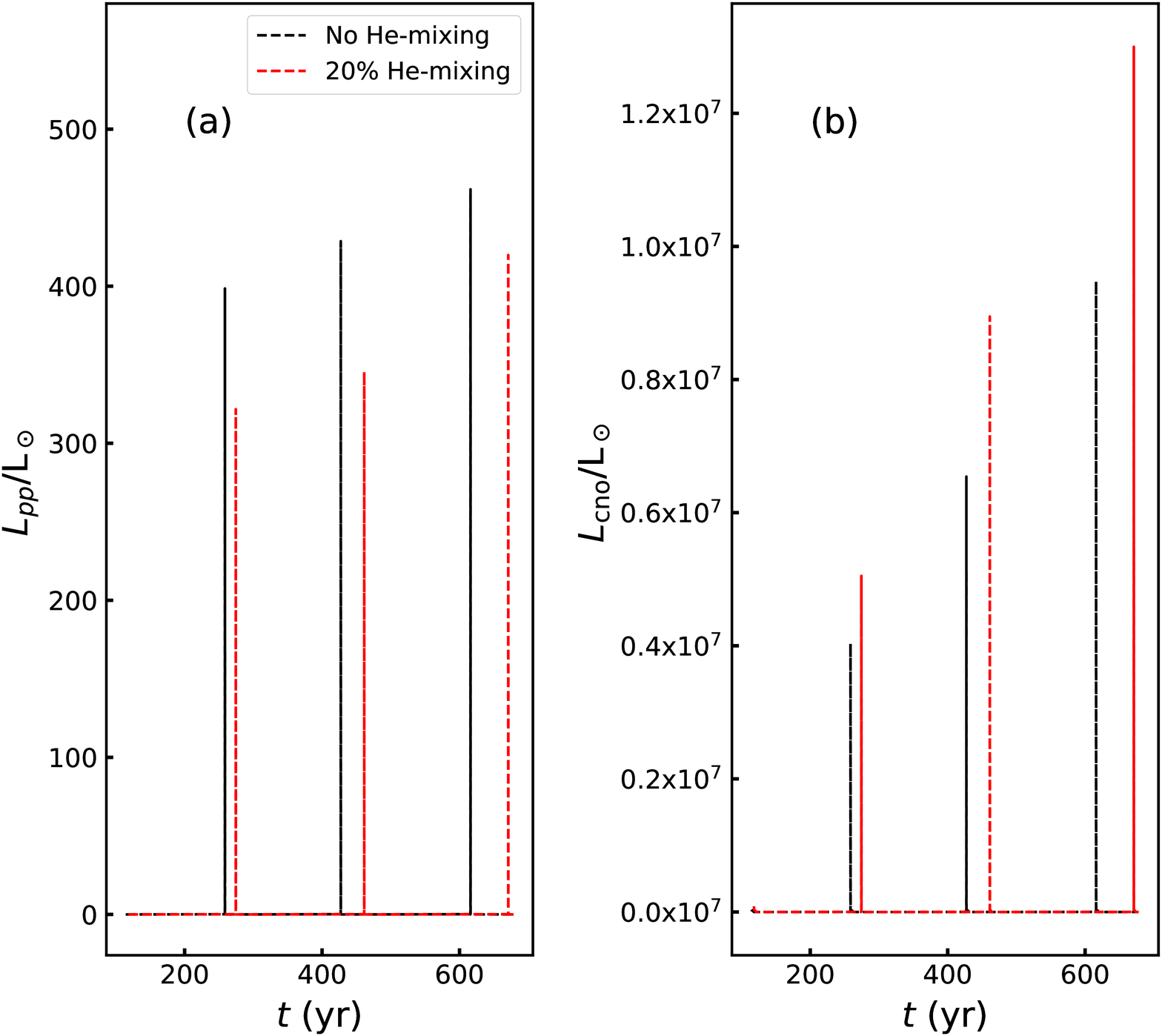}
	\caption{Comparison of the nuclear energy generation rate (expressed as a nuclear luminosity after multiplication by the mass of the envelope) of the $p$-$p$ chains (left panel) and CNO-cycle (right panel) between the He-mixing and no He-mixing model, in which we set the initial WD mass to be $1\,\rm M_\odot$.}
	\label{fig:pp}
\end{figure}

\begin{figure}
	\centering\includegraphics[width=\columnwidth*3/5]{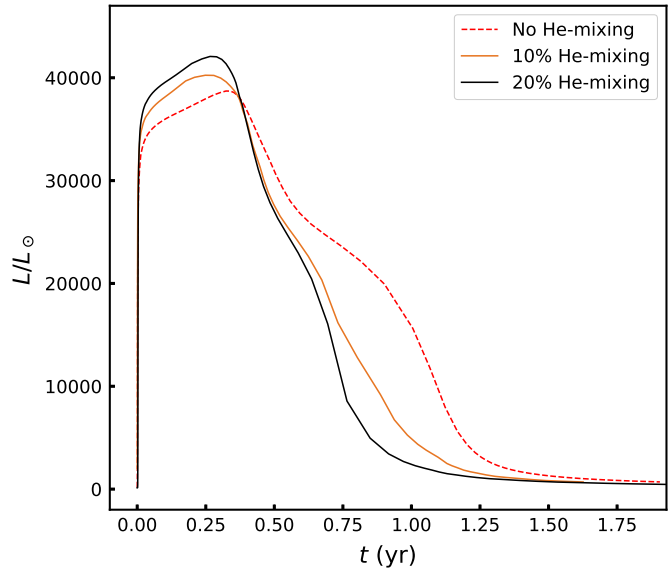}
	\caption{Comparison of luminosity evolution during a single nova outburst in different He-mixing levels, in which we set the initial WD mass to be $1.1\, \rm M_\odot$.}
	\label{fig:lum}
\end{figure}

\subsection{Comparison with observations}
\begin{table}
	\centering
	
\caption{Initial parameters of nova models.}	

	\begin{tabular}{ l  c c c ccc c c c  l }
		\toprule
		\hline \\
		Model&&	1 &2	&3	&4&5				  \\
		
		\hline \\
		$M_{\rm WD}$ $(\rm M_\odot)$&&0.70&0.70& 0.84&0.90 &1.05  \\
		$f_{\rm He}(\%)$&& 30&21&28&21&21\\
		$f_{\rm WD}(\%)$	&&20 &3&6 &10	&5   	   \\
		$\dot M_{\rm acc}$ ($\rm M_\odot /{\rm yr}$)&&5.0e-08&2.0e-10&1.8e-08&2.8e-09&9.0e-10	\\
		
		\hline
	\end{tabular}
\end{table}
We investigated the chemical abundances of five representative classical novae that show the enrichment of He in nova ejecta, i.e. GQ Mus, ASASSN-18fv, HR Del, T Aur and V443 Sct.
We note that these five novae also show significant enrichment of the WD material,
especially novae GQ Mus and T Aur.
Starrfield et al. (1998) suggested that the accreted material is first mixed with the He-shell by diffusion
or accretion-driven shear mixing,
and then mixed with the core material prior to the thermonuclear runaway.
In order to reproduce the observations,
both WD- and He-mixing are considered in our nova models.
We mixed the accreted material with a given mass fraction of the He-shell and WD material,
in which the composition of the WD material is assumed to be $X(^{12}\rm C)$ = 0.495, $X(^{16}\rm O)$ = 0.495, $X(^{22}\rm Ne)$ = 0.01.
Through a series of tests, we obtained the nova models that can be used to reproduce the observations.

The main properties of initial models used to reproduce these five novae are presented in Table 1, including $M_{\rm WD}$, $\dot M_{\rm acc}$, $f_{\rm He}$ and WD-mixing fraction ($f_{\rm WD}$).
$f_{\rm He}$ can be calculated by: $f_{\rm He} \times 0.72 = X_{\rm He}/(1-f_{\rm WD})-0.28$.
Table 2 lists the detailed nuclides in the nova ejecta of the five models.
The comparison between simulations and observations is shown in Table 3.
\begin{table}
	\centering
	\caption{Mass fraction of nuclides from nova models.}
	\begin{tabular}{ l  c  c ccc c  c  l }
		\toprule
		\hline \\
		Model&&1 &2 &3 &4 &5				  \\
		\hline \\
		$^{1}\rm H$&&3.73e-01&	5.13e-01&4.46e-01&4.89e-01&4.77e-01	\\
		$^{3}\rm He$&&7.06e-06& 4.44e-07&2.40e-06& 4.58e-06 &1.37e-07\\
		$^{4}\rm He$&&4.02e-01&4.42e-01&4.76e-01&4.45e-01&4.03e-01\\
		$^{7}\rm Li$&&7.25e-07& 1.58e-08&1.53e-07& 2.01e-07 &1.35e-08\\
		$^{12}\rm C$&&1.66e-02&7.84e-04&1.45e-03&1.63e-03&2.41e-03	\\
		$^{13}\rm C$&&9.02e-03&2.25e-04&3.82e-04&5.09e-04&7.66e-04	\\
		$^{14}\rm N$&&9.33e-02&2.47e-02&4.17e-02&3.93e-02&6.11e-02	\\
		$^{15}\rm N$&&3.93e-06&2.07e-05&2.01e-06&9.85e-06&2.04e-05	\\
		$^{16}\rm O$&&1.04e-01&1.58e-02&2.95e-02&1.94e-02&5.04e-02	\\
		$^{17}\rm O$&&3.35e-04&2.53e-04&1.28e-04&2.24e-04&4.32e-04	\\
		$^{18}\rm O$&&1.24e-05&6.02e-06&4.81e-06& 1.33e-05 &2.38e-06\\
		$^{19}\rm F$&&2.43e-07&7.09e-08&3.25e-08&5.32e-08&9.65e-08	\\
		$^{20}\rm Ne$&&1.34e-03&1.87e-03&1.87e-03&1.57e-03&1.50e-03	\\
		$^{21}\rm Ne$&&9.73e-08&2.80e-08&2.92e-08&2.60e-08& 3.53e-08\\
		$^{22}\rm Ne$&&2.11e-03&6.29e-04&6.25e-04&6.21e-04&1.12e-03	\\
		$^{22}\rm Na$&&3.38e-06&4.31e-06&3.36e-06& 3.81e-06& 1.74e-06\\
		$^{23}\rm Na$&&3.95e-04&3.51e-04&3.87e-04&8.01e-04& 1.09e-05\\
		$^{24}\rm Mg$&&4.23e-04&4.55e-05&5.42e-05&1.48e-06&9.27e-05	\\
		$^{25}\rm Mg$&&6.12e-05&5.70e-04&5.57e-04&6.01e-04&4.67e-04	\\
		$^{26}\rm Mg$&&6.42e-05&7.10e-05&5.51e-05&3.42e-05&6.85e-05	\\
		$^{26}\rm Al$&&9.94e-06&6.51e-06&1.27e-05&4.36e-05&4.74e-06	\\
		$^{27}\rm Al$&&5.02e-05&5.66e-05&5.74e-05&3.60e-05&5.54e-05	\\
		$^{28}\rm Si$&&5.66e-04&6.73e-04&6.70e-04&6.86e-04&6.37e-04	\\
		$^{29}\rm Si$&&2.97e-05&3.53e-05&3.51e-05&3.52e-05&3.34e-05	\\
		$^{30}\rm Si$&&2.03e-05&2.42e-05&2.40e-05&2.29e-05&2.55e-05\\
		$^{31}\rm P$&&5.64e-06&6.69e-06&6.65e-06&6.56e-06&6.34e-06	\\
		$^{32}\rm S$&&2.81e-04&3.33e-04&3.31e-04&3.33e-04&3.16e-04	\\
		$^{33}\rm S$&&2.29e-06&2.71e-06&2.70e-06&2.71e-06&2.57e-06	\\
		$^{34}\rm S$&&1.33e-05&1.57e-05&1.57e-05&1.57e-05&1.49e-05	\\
		$^{35}\rm Cl$&&3.01e-06&3.58e-06&3.56e-06&3.64e-06&3.38e-06	\\
		$^{36}\rm Ar$&&6.19e-05&7.35e-05&7.30e-05&7.33e-05&6.96e-05\\
		$^{37}\rm Cl$&&1.01e-06&1.20e-06&1.20e-06&1.20e-06&1.14e-06	\\
		$^{38}\rm Ar$&&1.19e-05&1.41e-05&1.41e-05&1.41e-05&1.34e-05	\\
		$^{39}\rm K$&&2.99e-06&3.56e-06&3.54e-06&3.37e-06 &3.60e-06\\
		$^{40}\rm Ca$&&5.13e-05&6.09e-05&6.05e-05&6.07e-05&5.77e-05	\\
		\hline
	\end{tabular}
\end{table}
\subsubsection{GQ Mus}
GQ Mus is one of the most well-observed classical novae detected by the X-ray satellite $ROSAT$ (e.g. {\"O}gelman et al. 1993; Orio, Covington \& {\"O}gelman 2001). 
Morisset \& Pequignot (1996) determined the chemical abundances for the ejecta of GQ Mus, including H to Fe. 
They found that the mass fraction of H and He in the ejecta are almost equal.
The abundances of C, N and O are 2.5, 100 and 10 times that of solar values, respectively.
Hachisu, Kato \& Cassatella (2008) suggested that the WD mass in GQ Mus is $0.7\,\rm M_\odot$.
According to the ejected mass, we set the mass-accretion rate to be $5 \times 10^{-8}\,\rm M_\odot \rm yr^{-1}$ (see Chen et al. 2019).
\begin{figure}
	\centering\includegraphics[width=\columnwidth*3/5]{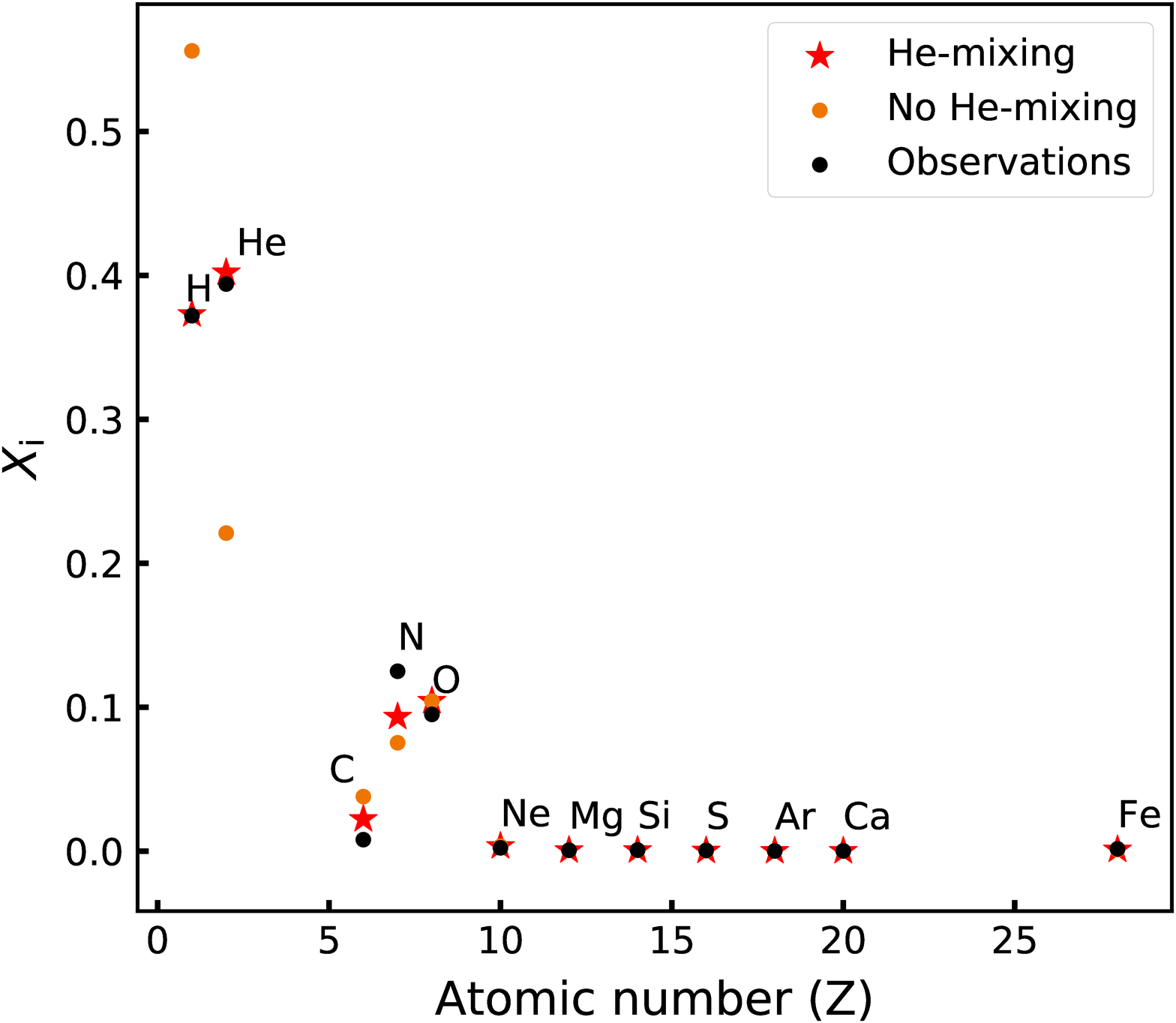}
	\caption{Comparison of the chemical abundances (mass fraction) in nova ejecta between the observations and our nova models, in which the black circles are the observed value obtained from Morisset \& Pequignot (1996), the red stars and orange circles are nova models with and without He-mixing, respectively. The WD material mixing is considered in the both nova models.}
	\label{fig:GQ}
\end{figure}
\begin{table*}
	\centering
	\caption{Models versus observations of some classical nova systems}
	\begin{tabular}{ lcccccccccl }
		\toprule
		\hline \\
		Model&&	H &He &$\sum \rm CNO$ &Ne &Na-Fe &Z &$\rm He/H$ \\
		\hline \\
		GQ Mus$^{1}$&& 0.37& 0.39& 0.23& 0.23e-02& 0.39e-02& 0.24& 1.05\\
		Model\,1&& 0.37& 0.40& 0.22& 0.35e-02& 0.26e-02& 0.23& 1.08\\
		Model\,1$^{\rm a}$&&0.56&0.22&0.21&0.35e-02&0.26e-02&0.22&0.60\\
		\hline \\
		ASASSN-18fv$^{2}$&&0.51&0.44&0.13e-01&...&...&0.50e-01&0.86\\
		Model\,2&&		0.51&	0.44&0.41e-01&	0.22e-02&	0.30e-02&	0.45e-01&0.80\\
		Model\,2$^{\rm a}$&&0.66&0.27&0.60e-01&0.22e-02&0.31e-02&0.70e-01&0.41\\
		\hline \\
		HR Del$^{3}$&&0.45&	0.48&0.74e-01&	0.30e-02&	...&		0.77e-01&	1.07\\
		Model\,3&&		0.45&	0.48&0.62e-01&		    0.22e-02&	0.31e-02&	0.77e-01&	1.07\\
		Model\,3$^{\rm a}$&&0.64&	0.30&0.63e-01&	0.22e-02&	0.32e-02&0.68e-01&0.47\\
		\hline \\
		T Aur$^{4}$&&	0.47&	0.40&0.13&...&		...&		0.13&	0.85\\
		Model\,4&&	0.48&	0.40&0.12&	0.26e-02&0.30e-02&	0.12&	0.85\\
		Model\,4$^{\rm a}$&&0.62&0.26&0.12&0.26e-02&0.30e-02&0.12&0.43\\
		\hline \\
		V443$^{5}$&&	0.49&	0.45&0.60e-01&0.14e-03&	0.17e-02&	0.62e-01&	0.92\\
		Model\,5&&		0.49&	0.45&0.60e-01&0.22e-02&	0.27e-02&	0.66e-01&	0.91\\	
		Model\,5$^{\rm a}$&&0.62&0.32&0.60e-01&0.22e-02&	0.31e-02&0.68e-01&0.51\\
		\hline
	\end{tabular}
	\begin{tablenotes}
		\item[]$^{\rm a}$ Nova model with WD-material mixing, but no He-mixing.
		\item[]References. 1: Morisset \& Pequignot (1996); 2: Pavana et al. (2020); 3: Tylenda (1978); 4: Gallagher et al. (1980); 5: Andrea et al. (1994).
	\end{tablenotes}
\end{table*}

Fig.\,6 shows the mass fraction of nova ejecta elements from our simulations and observations.
The nova model with $f_{\rm WD}=20\%$ and $f_{\rm He}=30\%$ can reproduce the observed abundances of nova ejecta well.
By comparing the nova models with or without He-mixing,
we can see that the abundance of H and He in the nova ejecta are sensitive to
whether the He-mixing is considered in nova models (see Table 3).
We note that the abundance of C and N in the observations is slightly different from our simulations.
This difference is caused by the composition of the mixed WD material selected in this work.
It should be noted that the He-mixing has almost no effect on the abundances of heavy elements for a given $f_{\rm WD}$ (see Table 3).
\subsubsection{ASASSN-18fv}
Pavana et al. (2020) studied the spectroscopic of the ejecta of ASASSN-18fv,
and suggested that the value of $\rm He/H$ and the metallicity Z are about 0.86 and 0.05, respectively.
The WD mass in ASASSN-18fv obtained by Pavana et al. (2020) is $0.7\,\rm M_\odot$.
The accretion rate is set to be $2 \times 10^{-10}\,\rm M_\odot \rm yr^{-1}$ (see Model 2 in Table 1).
The nova model with $21\%$ He-mixing fraction and $3\%$ WD-mixing fraction can reproduce the observations (see Table 3).
\subsubsection{HR Del}
The mass fraction of H and He in HR Del are 0.45 and 0.48, and the metallicity is 0.077 (see Tylenda 1978). 
Shara et al. (2018) suggested that the WD mass and accretion rate are $0.84\,\rm M_\odot$ and $1.86 \times 10^{-8}\,\rm M_\odot \rm yr^{-1}$, respectively (see Model 3 in Table 1).
Our simulations indicate that the He- and WD-mixing fraction of nova HR Del are $28\%$ and $6\%$, respectively (see Table 3).
\subsubsection{T Aur}
T Aur also shows high value of $\rm He/H$, but only H, He, N, O were determined in the observations (see Gallagher et al. 1980).
In this nova model, the WD mass and accretion rate are set to be $0.9\,\rm M_\odot$ and $2.8 \times 10^{-9}\,\rm M_\odot \rm yr^{-1}$, respectively (see Shara et al. 2018). 
The mixing fraction of WD material is 0.1 (see Model 4 in Table 1).
The abundance of H and He in our simulations are consistent with the observed values (see Table 3).
\subsubsection{V443}
The value of $\rm He/H$ in ejecta of V443 is about 0.92, and the metallicity Z is about 0.062 (see Andrea, Drechsel \& Starrfield 1994).
Shara et al. (2018) suggested that the WD mass and accretion rate are about $1.05\,\rm M_\odot$ and $9 \times 10^{-10}\,\rm M_\odot \rm yr^{-1}$, respectively.
The enrichment of He in the observations can be reproduced in the He-mixing model (see Table 3).
We note that the abundance of Ne in the observations is significantly lower than that in our simulations,
which may be due to the abundance of $^{22}\rm Ne$ of the WD in nova V443 is less than the 0.01 set in our models.
\subsection{He-mixing fractions}
\begin{figure}
	\centering\includegraphics[width=\columnwidth*3/5]{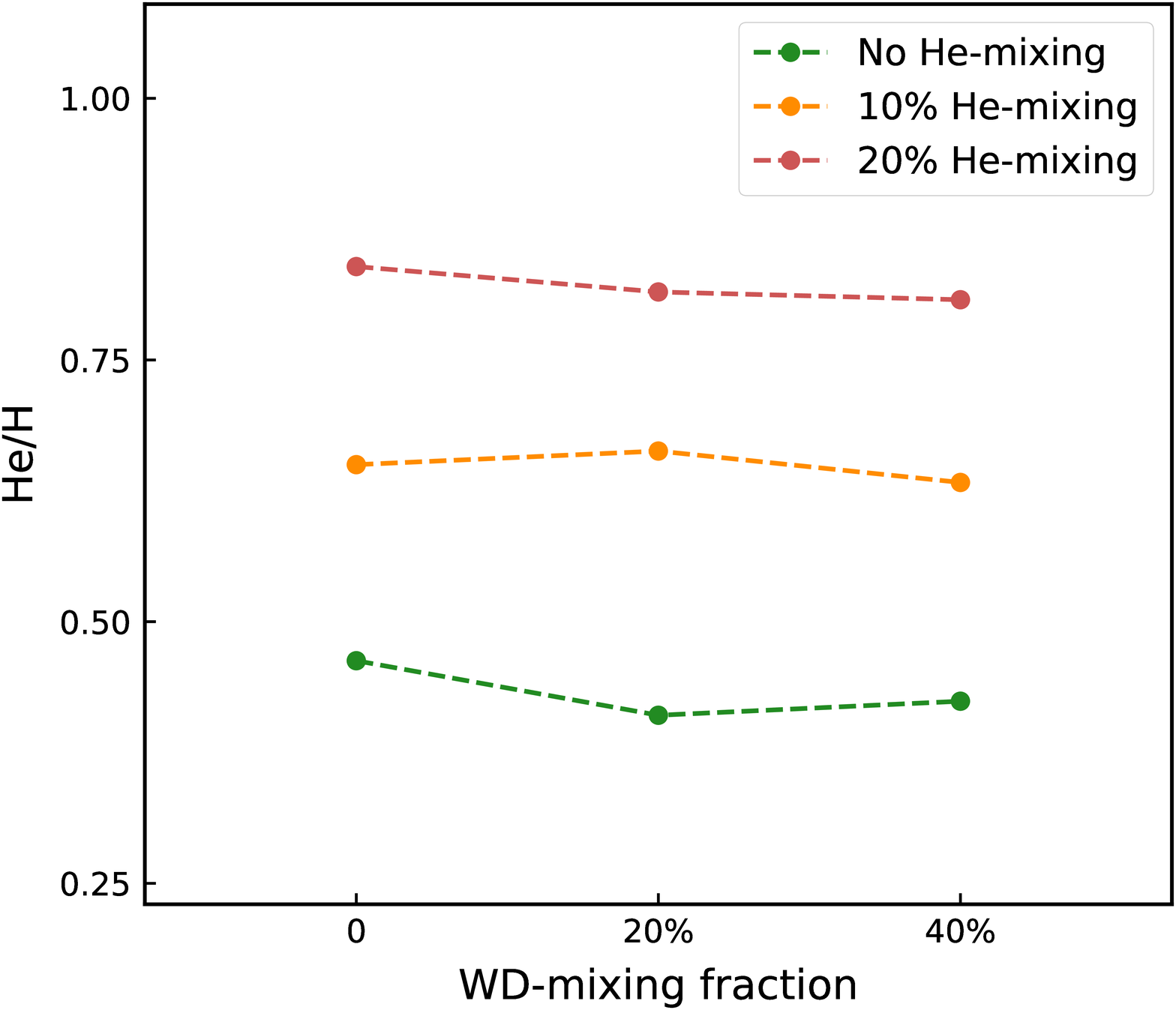}
	\caption{The changes of abundance ratio of He to H in the nova ejecta $\rm He/H$ with different He-mixing and WD-mixing levels, in which we set the initial WD mass to be $0.7\,\rm M_\odot$, and the WD-mixing level $\leq40\%$.}
	\label{fig:heding3}
\end{figure}

Kelly et al. (2013) suggested that the elemental abundance ratio that can be used to determine the mixing fraction should only be sensitive to the degree of mixing, but not sensitive to the WD mass.
As shown in Fig.\,2, $\rm He/H$ weakly depends on the WD mass ($M_{\rm WD}\leq1.1\,\rm M_\odot$),
but strongly depends on the He-mixing levels.
Meanwhile, we note that the He-mixing may occur simultaneously with the WD-mixing in some novae, e.g. GQ Mus and T Aur, etc.
The He-mixing meter should not be affected by the WD-mixing.
Fig.\,7 shows $\rm He/H$ in our simulations changing with the He-mixing fractions for various WD-mixing levels.
Obviously, $\rm He/H$ does not strongly depend on the WD-mixing levels.
Therefore, $\rm He/H$ can be used to estimate the He-mixing fraction in nova system with $M_{\rm WD}\leq1.1\,\rm M_\odot$. 

\begin{figure}
	\centering\includegraphics[width=\columnwidth*3/5]{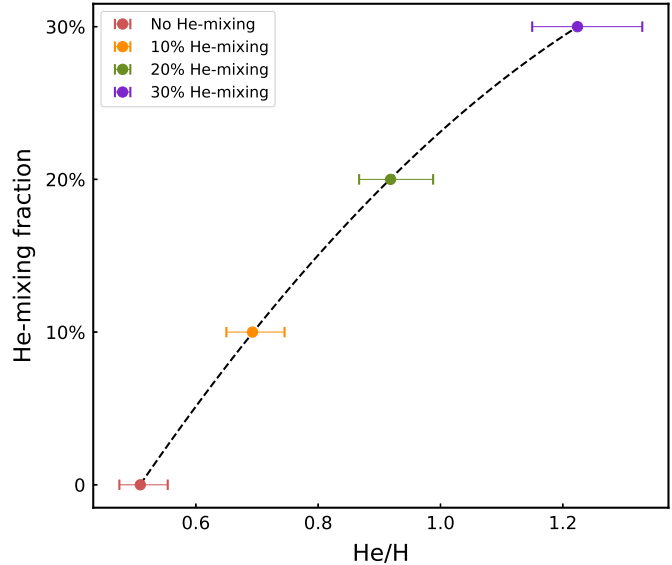}
	\caption{The value of $\rm He/H$ vs. He-mixing fraction, in which the error bar represents the effect of WD mass on $\rm He/H$. Different colors correspond to different He-mixing fractions. The dashed line gives the value of $f_{\rm He}$ expressed as a function of $\rm He/H$.  } 
	\label{fig:haibang1}
\end{figure}

\begin{table*}
	\centering
	
	\caption{He-mixing fraction of some classical novae estimated by $\rm He/H$.}	
	
	\begin{tabular}{ l  c c c ccc c c c  l }
		\toprule
		\hline \\
		Nova&&	RR Pic &V827 Her &V2214 Oph 	&DQ Her  &V842 Cen	 \\
		\hline
		$M_{\rm WD} (\rm M_\odot)$$^{\rm a}$&& 0.95    &1.10     &0.99      &0.95     &1.02\\
		$\rm He/H$$^{\rm b}$&&0.81    &0.81      &0.76		&0.59	&0.56\\
		$f_{\rm He} (\%)$&& 15      &15       &13     	&5        &3  \\
		\hline 
	\end{tabular}
	\begin{tablenotes}
		\item[]$^{\rm a}$ These WD masses are taken from Shara et al. (2018).
		\item[]$^{\rm b}$ The values of $\rm He/H$ in these novae are taken from Gehrz et al. (1998).
	\end{tablenotes}
\end{table*}

Fig.\,8 summaries the range of $\rm He/H$ in different degrees of He-mixing.
The error bar at each point represents the effect of the WD mass.
We used a polynomial function to fit the relationship between $f_{\rm He}$ and He/H,
and the fitted result can be expressed as follows:
\begin{equation}
f_{\rm He}=-0.23(X_{\rm He}/X_{\rm H})^2+0.816(X_{\rm He}/X_{\rm H})-0.356.
\end{equation}
By using the He-mixing meter $\rm He/H$,
we can estimate the He-mixing fraction in classical nova systems (see Table 4).

\section{Discussions and Summary}
\begin{figure}
	\centering\includegraphics[width=\columnwidth*3/5]{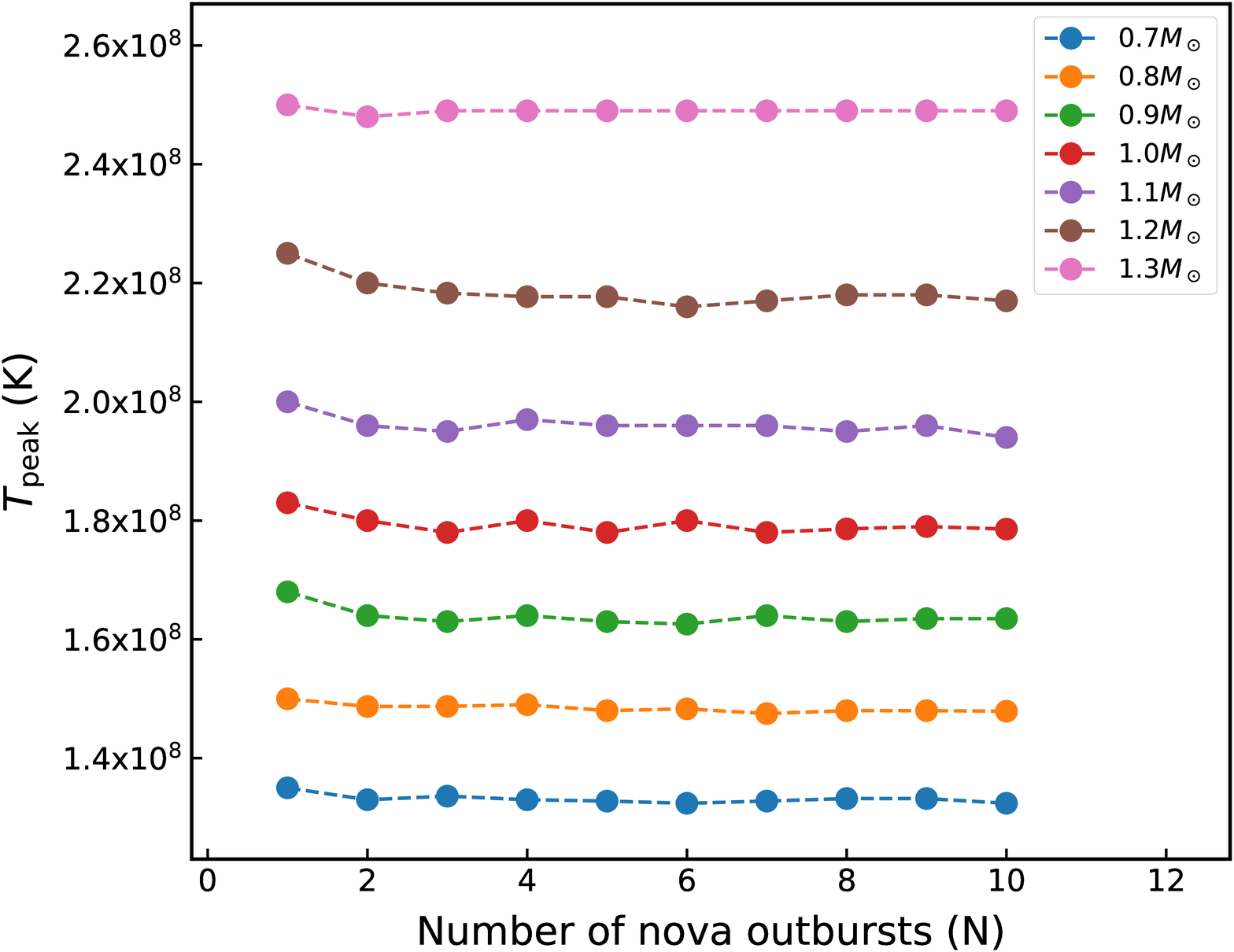}
	\caption{The peak temperature during every single nova outburst, in which we set the initial WD masses in the range of $0.7-1.3\,\rm M_\odot$, and the accretion rate to be $2 \times 10^{-9}\,\rm M_\odot \rm yr^{-1}$.}
	\label{fig:guai2}
\end{figure}
In the present work, we studied the variation of $\rm He/H$ with the WD mass for different He-mixing fractions.
Our simulations can roughly estimate whether the He-mixing process has occurred between
the accreted H-rich matter and the He-shell.
We found that the value of He/H would be about 0.5 in no He-mixing model,
and can exceed 1 if the He-mixing fraction is assumed as $30\%$ ($M_{\rm WD}\leq1.1\,\rm M_\odot$).
In addition, $\rm He/H$ is sensitive to the WD mass, which increases obviously if the WD mass is greater than $1.1\,\rm M_\odot$.
In order to explain this phenomenon,
we performed a series of calculations on the nova model without He-/WD-mixing.
Fig.\,9 represents the peak temperature during nova outbursts changing with the WD mass,
where we set the mass accretion rate to be $2 \times 10^{-9}\,\rm M_\odot \rm yr^{-1}$.
Obviously, the peak temperature increases more rapidly when the WD mass is greater than $1.1\,\rm M_\odot$,
causing more H to burn into He.
Meanwhile, it is worth noting that $D$ and $M_{\rm ej}$ also decrease with the WD mass significantly when the WD mass is greater than $1.1\,\rm M_\odot$ (see Fig.\,3).

We provide a fitting formula to estimate the He-mixing fraction in classical novae by the value of He/H (see Equation 3).
Note that this fitting formula may not apply to ONe novae.
The reason is that the mass of ONe WDs is usually greater than $1.15\,\rm M_\odot$,
and WDs in this mass range will cause He/H to change more rapidly,
thereby reducing the accuracy of measuring He-mixing fraction.
On the other hand, 
it has been suggested that an ONe WD has a CO-buffer layer and a He-shell (e.g. Gil-Pons \& Garc{\'\i}a-Berro 2001; Gil-Pons et al. 2003).
Before significant amounts of Ne can be detected,
the CO buffer and the He-shell need to be continuously dredged up and blown off.
This means that the He-shell would not be accumulated on the surface of ONe WDs,
and thus the He-enrichment may not occur in ONe novae.

 
By adopting the so-called pre-mixed model, our simulations can reproduce the chemical abundances in five representative novae.
However, the multi-dimensional studies suggested that the obvious mixing process will only occur when a strong TNR condition is about to be reached (e.g. Glasner, Livne \& Truran 1997, 2012).
Jos{\'e}, Shore \& Casanova (2020) explored a new methodology that combines 1D and 3D simulations to study nova outbursts,
known as the $123$-$321$ nova model.
Their model can eliminate the mixing fraction parameter required in the pre-mixed model.
They compared the results of their simulations with the pre-mixed model.
There are some differences in the abundance of Mg-Al,
and the abundance of $^{7}\rm Li$ in their results is much smaller than that in the pre-mixed model.
However, the abundance of other chemical elements and the metallicity in nova ejecta are almost unchanged for CO WD material mixing.
Therefore, the pre-mixed model in the present work is appropriate.

In this work, we calculated a large number of accreting WDs by assuming the metallicity of the nova population as 0.02.
However, it has been suggested that novae with different populations show various characteristics due to the effect of the metallicities.
Wu et al. (2017) found that the nova cycle duration would decrease in a higher metallicity population.
Chen et al. (2019) studied the influence of population on nova outbursts by assuming different metallicities.
They found that the accreted and ejected mass decrease with the metallicities,
and that the nova outbursts would be more violent in the low metallicities (see also Jos{\'e} et al. 2007).
The mass fraction of He in the donor can be calculated by $Y=0.24+2\rm Z$,
where Z represents the metallicity.
According to this equation,
a lower initial He abundance in the accreted material would be expected in the low metallicity populations,
which may cause the value of $\rm He/H$ to be decreased.

By using the stellar evolution code MESA, we investigated the influence of He-mixing on the properties of nova outbursts.
We found that the mass fraction of He and H in nova ejecta are sensitive to the He-mixing levels.
Meanwhile, the WD mass can affect the abundance of He and H significantly if its mass is larger than $1.1\rm M_\odot$.
We also found that the nova cycle duration and ejected mass would increase if the accreted material is mixed with the He-shell.
In addition, the nuclear energy generation rate of $p$-$p$ chains decreases with the He-mixing level during nova outbursts,
whereas the CNO-cycle increases.
The chemical abundances of five novae (e.g. GQ Mus, ASASSN-18fv, HR Del, T Aur and V443) can be reproduced well in the present work.
Our simulations imply that both He and WD material mixing need to be taken into account when studying nova outbursts. 
Moreover, this work develops a He-mixing meter (i.e. He/H)
that can be used to estimate the He-mixing fraction in nova systems.
In order to better understand the nova outbursts,
more theoretical studies are required,
and large nova samples are needed in the observations.
 
\section*{Acknowledgements}
We thank the referee Jordi Jos{\'e} for valuable comments that help to improve the paper.
BW is supported by the National Natural Science Foundation of China (Nos 11873085 and 11521303),
the Chinese Academy of Sciences (No QYZDB-SSW-SYS001),
and the Yunnan Province (Nos 2018FB005, 2019FJ001 and 202001AS070029).
CW is supported by the National Natural Science Foundation of China (No. 12003013).
This work is also supported by the CSST project of ``stellar activity and later stage of evolution'',
and the National Natural Science Foundation of China (No. U2031116).


\begin{thebibliography}{}
\bibitem[Alexakis et al. (2004)]{alex04}			Alexakis A. et al., 2004, ApJ, 602, 931
\bibitem[Andrea, Drechsel, \& Starrfield (1994)]{andr94}  Andrea J., Drechsel H., Starrfield S., 1994, A\&A, 291, 869
\bibitem[Casanova et al. (2010)]{casa10}            Casanova J., Jos{\'e} J., Garc{\'\i}a-Berro E., Calder A., Shore S.~N., 2010, A\&A, 513, L5
\bibitem[Casanova et al. (2016)]{casa16}            Casanova J., Jos{\'e} J., Garc{\'\i}a-Berro E., Shore S.~N., 2016, A\&A, 595, A28
\bibitem[Chen et al. (2019)]{chen19}                Chen H.-L., Woods T.~E., Yungelson L.~R., Piersanti L., Gilfanov M., Han Z., 2019, MNRAS, 490, 1678
\bibitem[Denissenkov et al. (2013)]{deni13}         Denissenkov P.~A., Herwig F., Bildsten L., Paxton B., 2013, ApJ, 762, 8
\bibitem[Denissenkov et al. (2014)]{deni14}         Denissenkov P.~A. et al., 2014, MNRAS, 442, 2058
\bibitem[Denissenkov et al. (2017)]{deni17}         Denissenkov P.~A., Herwig F., Battino U., Ritter C., Pignatari M., Jones S., Paxton B., 2017, ApJL, 834, L10
\bibitem[Downen et al. (2013)]{down13}              Downen L.~N., Iliadis C., Jos{\'e} J., Starrfield S., 2013, ApJ, 762, 105
\bibitem[Durisen (1977)]{duri97}              		Durisen R.~H., 1977, ApJ, 213, 145
\bibitem[Evans et al. (2001)]{evan01}              	Evans A., Krautter J., Vanzi L., Starrfield S., 2001, A\&A, 378, 132
\bibitem[Fujimoto (1988)]{fuji88}              		Fujimoto M.~Y., 1988, A\&A, 198, 163
\bibitem[Fujimoto \& Iben (1992)]{fuji92}           Fujimoto M., Iben I., 1992, ApJ, 399, 646
\bibitem[Gallagher et al. (1980)]{gall80}           Gallagher J.~S., Hege E.~K., Kopriva D.~A., Williams R.~E., Butcher H.~R., 1980, ApJ, 237, 55
\bibitem[Gehrz et al. (1998)]{gehr98}           	Gehrz R.~D., Truran J.~W., Williams R.~E., Starrfield S., 1998, PASP, 110, 3
\bibitem[Gil-Pons \& Garc{\'\i}a-Berro (2001)]{gilp01}      Gil-Pons P., Garc{\'\i}a-Berro E., 2001, A\&A, 375, 87
\bibitem[Gil-Pons et al. (2003)]{gilp03}           	Gil-Pons P., Garc{\'\i}a-Berro E., Jos{\'e} J., Hernanz M., Truran J.~W., 2003, A\&A, 407, 1021
\bibitem[Glasner \& Livne (1995)]{glas95}           Glasner S.~A., Livne E., 1995, ApJL, 445, L149
\bibitem[Glasner, Livne, \& Truran (1997)]{glas97}  Glasner S.~A., Livne E., Truran J.~W., 1997, ApJ, 475, 754
\bibitem[Glasner, Livne, \& Truran (2012)]{glas12}  Glasner S.~A., Livne E., Truran J.~W., 2012, MNRAS, 427, 2411
\bibitem[Guo et al. (2021)]{guo21}  				Guo Y., Liu D., Wu C., Wang B., 2021, RAA, 21, 034
\bibitem[Hachisu, Kato, \& Cassatella (2008)]{hach08}  Hachisu I., Kato M., Cassatella A., 2008, ApJ, 687, 1236
\bibitem[Hillebrandt \& Niemeyer (2000)]{hill00}  	Hillebrandt W., Niemeyer J.~C., 2000, ARA\&A, 38, 191
\bibitem[Hillman et al. (2016)]{hill16}  			Hillman Y., Prialnik D., Kovetz A., Shara M.~M., 2016, ApJ, 819, 168
\bibitem[Idan, Shaviv, \& Shaviv (2013)]{idan13}  	Idan I., Shaviv N.~J., Shaviv G., 2013, MNRAS, 433, 2884
\bibitem[Iglesias \& Rogers (1993)]{igle93}  		Iglesias C.~A., Rogers F.~J., 1993, ApJ, 412, 752
\bibitem[Iglesias \& Rogers (1996)]{igle96}  		Iglesias C.~A., Rogers F.~J., 1996, ApJ, 464, 943
\bibitem[Jose \& Hernanz (1998)]{jose98}  			Jos{\'e} J., Hernanz M., 1998, ApJ, 494, 680
\bibitem[Jose et al. (2007)]{jose07}  				Jos{\'e} J., Garc{\'\i}a-Berro E., Hernanz M., Gil-Pons P., 2007, ApJL, 662, L103
\bibitem[Jose (2016)]{jose16}  						Jos{\'e} J., 2016, Stellar Explosions: Hydrodynamics and Nucleosynthesis, RC
Press/Taylor and Francis, Boca Raton, FL, p. 147
\bibitem[Jose, Shore, \& Casanova (2020)]{jose20}  	Jos{\'e} J., Shore S.~N., Casanova J., 2020, A\&A, 634, A5
\bibitem[Kelly et al. (2013)]{kell13}  				Kelly K.~J., Iliadis C., Downen L., Jos{\'e} J., Champagne A., 2013, ApJ, 777, 130
\bibitem[Kippenhahn \& Thomas (1978)]{kipp78}  		Kippenhahn R., Thomas H.-C., 1978, A\&A, 63, 265
\bibitem[Kovetz \& Prialnik (1997)]{kove97}  		Kovetz A., Prialnik D., 1997, ApJ, 477, 356
\bibitem[Kutter \& Sparks (1987)]{kutt87}  			Kutter G.~S., Sparks W.~M., 1987, ApJ, 321, 386
\bibitem[Livio \& Truran (1994)]{livi94}  			Livio M., Truran J.~W., 1994, ApJ, 425, 797
\bibitem[MacDonald (1983)]{macd83}  				MacDonald J., 1983, ApJ, 273, 289
\bibitem[MacDonald (1984)]{macd84}  				MacDonald J., 1984, ApJ, 283, 241
\bibitem[Morisset \& Pequignot (1996)]{mori96}  	Morisset C., Pequignot D., 1996, A\&A, 312, 135
\bibitem[Nomoto et al. (2007)]{nomo07}  			Nomoto K., Saio H., Kato M., Hachisu I., 2007, ApJ, 663, 1269
\bibitem[Oegelman et al. (1993)]{oege93}  	{\"O}egelman H., Orio M., Krautter J., Starrfield S., 1993, Natur, 361, 331
\bibitem[Orio, Covington, \& {\"O}gelman (2001)]{orio01}  	Orio M., Covington J., {\"O}gelman H., 2001, A\&A, 373, 542
\bibitem[Pavana et al. (2020)]{pava20}  	Pavana M., Raj A., Bohlsen T., Anupama G.~C., Gupta R., Selvakumar G., 2020, MNRAS, 495, 2075
\bibitem[Paxton et al. (2011)]{paxt11}  	Paxton B., Bildsten L., Dotter A., Herwig F., Lesaffre P., Timmes F., 2011, ApJS, 192, 3
\bibitem[Paxton et al. (2013)]{paxt13}  	Paxton B. et al., 2013, ApJS, 208, 4
\bibitem[Paxton et al. (2015)]{paxt15}  	Paxton B. et al., 2015, ApJS, 220, 15
\bibitem[Paxton et al. (2018)]{paxt18}  	Paxton B. et al., 2018, ApJS, 234, 34
\bibitem[Politano et al. (1995)]{poli95}  	Politano M., Starrfield S., Truran J.~W., Weiss A., Sparks W.~M., 1995, ApJ, 448, 807
\bibitem[Prialnik \& Kovetz (1984)]{pria84} Prialnik D., Kovetz A., 1984, ApJ, 281, 367
\bibitem[Rosner et al. (2001)]{rosn01} 		Rosner R., Alexakis A., Young Y.-N., Truran J.~W., Hillebrandt W., 2001, ApJL, 562, L177
\bibitem[Rukeya et al. (2017)]{ruke17} 		Rukeya R., L{\"u} G., Wang Z., Zhu C., 2017, PASP, 129, 074201
\bibitem[Shafter (2017)]{shaf17} 			Shafter A.~W., 2017, ApJ, 834, 196
\bibitem[Shara et al. (2018)]{shar18} 		Shara M.~M., Prialnik D., Hillman Y., Kovetz A., 2018, ApJ, 860, 110
\bibitem[Starrfield et al. (1972)]{star72} 	Starrfield S., Truran J.~W., Sparks W.~M., Kutter G.~S., 1972, ApJ, 176, 169
\bibitem[Starrfield et al. (1998)]{star98} 	Starrfield S., Truran J.~W., Wiescher M.~C., Sparks W.~M., 1998, MNRAS, 296, 502
\bibitem[Starrfield, Iliadis, \& Hix (2008)]{star08} 	Starrfield S., Iliadis C., Hix W. R., 2008, in Bode M. F., Evans A., eds, 2nd edn. Cambridge Univ. Press, Cambridge, UK, p. 77
\bibitem[Starrfield, Iliadis, \& Hix (2016)]{star16} 	Starrfield S., Iliadis C., Hix W.~R., 2016, PASP, 128, 051001
\bibitem[Starrfield et al. (2020)]{star20} 	Starrfield S., Bose M., Iliadis C., Hix W.~R., Woodward C.~E., Wagner R.~M., 2020, ApJ, 895, 70
\bibitem[Tauris et al. (2013)]{taur13} 		Tauris T.~M., Sanyal D., Yoon S.-C., Langer N., 2013, A\&A, 558, A39
\bibitem[Truran \& Livio (1986)]{trur86} 	Truran J.~W., Livio M., 1986, ApJ, 308, 721
\bibitem[Tylenda (1978)]{tyle78} 			Tylenda R., 1978, AcA, 28, 333
\bibitem[Wang \& Han (2012)]{wang12} 		Wang B., Han Z., 2012, NewAR, 56, 122
\bibitem[Wang (2018)]{wang18} 				Wang B., 2018, RAA, 18, 049
\bibitem[Wang \& Liu (2020)]{wang20} 		Wang B., Liu D., 2020, RAA, 20, 135
\bibitem[Webbink et al. (1987)]{webb87} 	Webbink R.~F., Livio M., Truran J.~W., Orio M., 1987, ApJ, 314, 653
\bibitem[Wolf et al. (2013)]{wolf13} 		Wolf W.~M., Bildsten L., Brooks J., Paxton B., 2013, ApJ, 777, 136
\bibitem[Woosley (1986)]{woos86} 			Woosley S.~E., 1986, nce..conf, 1
\bibitem[Wu et al. (2017)]{wu17} 			Wu C., Wang B., Liu D., Han Z., 2017, A\&A, 604, A31
\bibitem[Yaron et al. (2005)]{yaro05} 		Yaron O., Prialnik D., Shara M.~M., Kovetz A., 2005, ApJ, 623, 398
\end{thebibliography}


\section*{Data availability}
Results will be shared on reasonable request to corresponding author.


\label{lastpage}
\end{document}